\documentclass[%
 reprint,
 twocolumn,
 superscriptaddress,
 amsmath,amssymb,
 aps,
]{revtex4-2}

\usepackage{graphicx}
\usepackage{dcolumn}
\usepackage{bm}
\usepackage{color} 

\begin{document}


\title{Analytical approach to the generalized friendship paradox in networks with correlated attributes}%

\author{Hang-Hyun Jo}
\email{h2jo@catholic.ac.kr}
\affiliation{%
 Department of Physics, The Catholic University of Korea, Bucheon 14662, Republic of Korea
}%

\author{Eun Lee}
\affiliation{Department of Computer Science, University of Colorado Boulder, Boulder, Colorado 80309, USA}
\affiliation{BioFrontiers Institute, University of Colorado Boulder, Boulder, Colorado 80309, USA}

\author{Young-Ho Eom}
\affiliation{Department of Physics, University of Seoul, Seoul 02504, Republic of Korea}
\affiliation{Urban Big data and AI Institute, University of Seoul, Seoul 02504, Republic of Korea}

\date{\today}

\begin{abstract}
One of the interesting phenomena due to the topological heterogeneities in complex networks is the friendship paradox, stating that your friends have on average more friends than you do. Recently, this paradox has been generalized for arbitrary nodal attributes, called a generalized friendship paradox (GFP). In this paper, we analyze the GFP for the networks in which the attributes of neighboring nodes are correlated with each other. The correlation structure between attributes of neighboring nodes is modeled by the Farlie-Gumbel-Morgenstern copula, enabling us to derive approximate analytical solutions of the GFP for three kinds of methods summarizing the neighborhood of the focal node, i.e., mean-based, median-based, and fraction-based methods. The analytical solutions are comparable to simulation results, while some systematic deviations between them might be attributed to the higher-order correlations between nodal attributes. These results help us get deeper insight into how various summarization methods as well as the correlation structure of nodal attributes affect the GFP behavior, hence better understand various related phenomena in complex networks.
\end{abstract}

\maketitle


\section{Introduction}\label{sec:intro}

In recent years, complex social phenomena have been extensively studied by means of statistical physics~\cite{Castellano2009Statistical, Sen2014Sociophysics} due to the structural similarity between social phenomena and physical processes in a sense that macroscopic complex patterns emerge from the interaction between numerous microscopic constituents. The interaction structure between individuals has been modeled in terms of social networks, where nodes and links represent individuals and their pairwise interactions, respectively~\cite{Albert2002Statistical, Borgatti2009Network, Menczer2020First}. Empirical analyses of various social network datasets have revealed that the topological structure of social networks is heterogeneous~\cite{Jo2018Stylized}, typically showing heavy-tailed degree distributions~\cite{Barabasi1999Emergence, Broido2019Scalefree}, assortative mixing~\cite{Newman2002Assortative}, and community structure~\cite{Fortunato2010Community}, etc. Such heterogeneous topological properties enable various interesting phenomena in social networks and social processes taking place on them, such as diffusion, spreading, and opinion formation~\cite{Castellano2009Statistical, Sen2014Sociophysics, Pastor-Satorras2015Epidemic, Masuda2017Random}.

Among various interesting phenomena due to the topological heterogeneities of social networks we focus on the friendship paradox (FP) and its generalized version, namely, the generalized friendship paradox (GFP). The FP states that your friends have on average more friends than you do~\cite{Feld1991Why}. In terms of network science, the FP is about the node degree, i.e., the number of neighbors of the node. On the other hand, the GFP can be applied to the network of nodes having attributes other than degrees whether such attributes are topological, e.g., betweenness or eigenvector centralities~\cite{Grund2014Why, Higham2019Centralityfriendship}, or non-topological, e.g., happiness or sentiment~\cite{Bollen2017Happiness, Zhou2020Sentiment}: The GFP states that your friends have on average higher attributes than yours~\cite{Hodas2013Friendship, Eom2014Generalized, Jo2014Generalized}. Since its introduction, the GFP has been extensively studied by means of empirical analyses~\cite{Hodas2013Friendship, Eom2014Generalized, Lerman2016Majority, Momeni2016Qualities, Benevenuto2016Hindex, Bollen2017Happiness, Alipourfard2020Friendship, Zhou2020Sentiment} as well as by analytical and numerical approaches~\cite{Jo2014Generalized, Fotouhi2015Generalized, Higham2019Centralityfriendship}. 

Both FP and GFP are based on the comparison of one node's attribute to a set of attributes of its neighbors or a single value summarizing the set. The most common summarization method has been to take an average of the attributes in the set, which is however sensitive on a few neighbors with very high attributes. Therefore, the median was suggested for the summarization as the median is less sensitive on such neighbors than the average~\cite{Feld1991Why, Ugander2011Anatomy, Kooti2014Network, Momeni2016Qualities}. More recently, another summarization method using the fraction of neighbors having higher attributes than the node of interest has been suggested to systematically compare different summarization methods~\cite{Lee2019Impact}. These three summarization methods are called mean-based, median-based, and fraction-based, respectively. Each summarization method can be interpreted as a perception model by which individuals perceive their neighborhood. It has been shown that different summarization methods lead to qualitatively different behaviors of FP and opinion formation~\cite{Lee2019Impact}.

As a natural extension of the previous work on the effect of summarization methods on the FP, here we study how the above three summarization methods affect the GFP behavior. For more rigorous understanding of such effects, we derive approximate analytical solutions of the probability that the GFP holds for a node with given degree and non-topological attribute, for each of three summarization methods. We interpret the GFP holding probability as the peer pressure on the individual node. It is straightforward to get the solution for the uncorrelated case, i.e., when the attributes of neighboring nodes are uncorrelated, e.g., as shown in Ref.~\cite{Jo2014Generalized}. In contrast, it has been highly non-trivial to derive the analytical solution for the case with correlated attributes, mainly due to the difficulty in modeling the correlation structure between attributes of neighboring nodes in a closed form. For modeling the correlated attributes of neighboring nodes, we adopt a Farlie-Gumbel-Morgenstern copula among others~\cite{Nelsen2006Introduction, Cossette2013Multivariate} to derive approximate analytical solutions of the mean-based, fraction-based, and median-based peer pressure for the first time to the best of our knowledge. The analytical solutions are compared to the simulation results. These results can help us get deeper insight into how various summarization methods as well as the correlation structure of nodal attributes affect the GFP behavior, hence better understand various related phenomena in complex networks.

\section{Mean-based GFP}\label{sec:mean}

Let us consider a network with $N$ nodes and a non-topological attribute distribution $P(x)$ for nodes. We assume the range of $x\geq 0$ for the sake of convenience, while the negative value of attributes can also be considered. The generalized friendship paradox (GFP) at the individual level is based on the comparison of one node's attribute to a set of attributes of its neighbors or a single value summarizing the set. We consider three different summarization methods, which are mean-based, median-based, and fraction-based, respectively~\cite{Lee2019Impact}.

\subsection{Analysis}\label{subsec:mean_anal}

The mean-based GFP holds for a node $i$ if the node has lower attribute than the average attribute of its neighbors, precisely if the following condition is satisfied~\cite{Eom2014Generalized, Jo2014Generalized}:
\begin{align}
  \frac{1}{k_i}\sum_{j\in\Lambda_i}x_j > x_i,
  \label{eq:mean_define_original}
\end{align}
where $\Lambda_i$ denotes the set of $i$'s neighbors and $k_i\equiv |\Lambda_i|$ is the degree of the node $i$. The probability of satisfying Eq.~(\ref{eq:mean_define_original}) is called the mean-based peer pressure. The mean-based peer pressure of a focal node with degree $k$ and attribute $x$ can be written as
\begin{align}
  & h_{\rm mn}(k,x) \equiv \Pr\left(\frac{1}{k}\sum_{j=1}^k x_j>x\right) = \Bigg\langle \theta\left(\frac{1}{k}\sum_{j=1}^k x_j-x\right)\Bigg\rangle \nonumber \\
  & =\prod_{j=1}^k \int_0^\infty dx_j P(x_1,\cdots,x_k|x)\theta\left(\frac{1}{k}\sum_{j=1}^k x_j-x\right),
  \label{eq:mean_define}
\end{align}
where $\theta(\cdot)$ is a Heaviside step function, $\langle\cdot\rangle$ denotes the ensemble average over $\{x_j\}$, and $P(x_1,\cdots,x_k|x)$ is the conditional joint probability distribution function (PDF) of $k$ attributes of neighbors of the focal node when the attribute of the focal node is given as $x$. For the analysis, we assume that $x_j$s are independent of each other but only conditioned by $x$, which is called the conditional independence. By this assumption the conditional joint PDF of $k$ attributes reduces to the product of $k$ conditional PDFs as follows:
\begin{align}
  P(x_1,\cdots,x_k|x)=\prod_{j=1}^k P(x_j|x),\ P(x_j|x)=\frac{P(x_j,x)}{P(x)}.
  \label{eq:joint_Px}
\end{align}
Here the joint PDF $P(x,x')$ carries information on the pairwise correlation between attributes of neighboring nodes. 

For modeling $P(x,x')$, we adopt a Farlie-Gumbel-Morgenstern (FGM) copula among others~\cite{Nelsen2006Introduction, Cossette2013Multivariate}, enabling us to write
\begin{align}
  P(x,x')=P(x)P(x')[1+rf(x)f(x')],
  \label{eq:Pxx}
\end{align}
where 
\begin{align}
    f(x)\equiv 2F(x)-1,\ F(x)\equiv \int_0^x P(y)dy.
\end{align}
The FGM copula indicates a function $C$ joining a bivariate cumulative distribution function (CDF), say $G(x,y)$, to their one-dimensional marginal CDFs, say $u(x)$ and $v(y)$, such that $G(x,y) = C[u(x), v(y)] = uv[1+r(1-u)(1-v)]$~\cite{Nelsen2006Introduction, Cossette2013Multivariate}. Then the bivariate PDF of $x$ and $y$ is derived by $\frac{\partial^2 G}{\partial x\partial y} = P_1(x)P_2(y)[1+r(2u-1)(2v-1)]$, where $P_1(x)$ and $P_2(y)$ denote PDFs. In our case, $P_1$ and $P_2$ are identical and the parameter $r\in [-1,1]$ controls the degree of correlations between $x$ and $x'$ in Eq.~\eqref{eq:Pxx}. Thus, $r$ is related to the Pearson correlation coefficient between $x$ and $x'$, which is written as
\begin{align}
    \rho_{x}\equiv \frac{\langle xx'\rangle -\mu^2}{\sigma^2},
\end{align}
where 
\begin{align}
    \langle xx'\rangle\equiv \int_0^\infty dx \int_0^\infty dx' xx' P(x,x'),
\end{align}
and $\mu$ and $\sigma$ are the mean and standard deviation of $P(x)$, respectively. Using Eq.~\eqref{eq:Pxx} one gets
\begin{align}
    \rho_{x}=\frac{r}{\sigma^2}\left[\int_0^\infty dx x P(x)f(x)\right]^2\equiv Ar.
    \label{eq:rhoxx_r}
\end{align}
The upper bound of $A$ for any $P(x)$ is known as $1/3$, implying that $|\rho_{x}|\le 1/3$~\cite{Schucany1978Correlation}. The FGM copula has recently been used for modeling the bivariate luminosity function of galaxies~\cite{Takeuchi2010Constructing} and bursty time series with correlated interevent times~\cite{Jo2019Analytically, Jo2019Copulabased}.

\begin{figure*}[!t]
\includegraphics[width=0.9\textwidth]{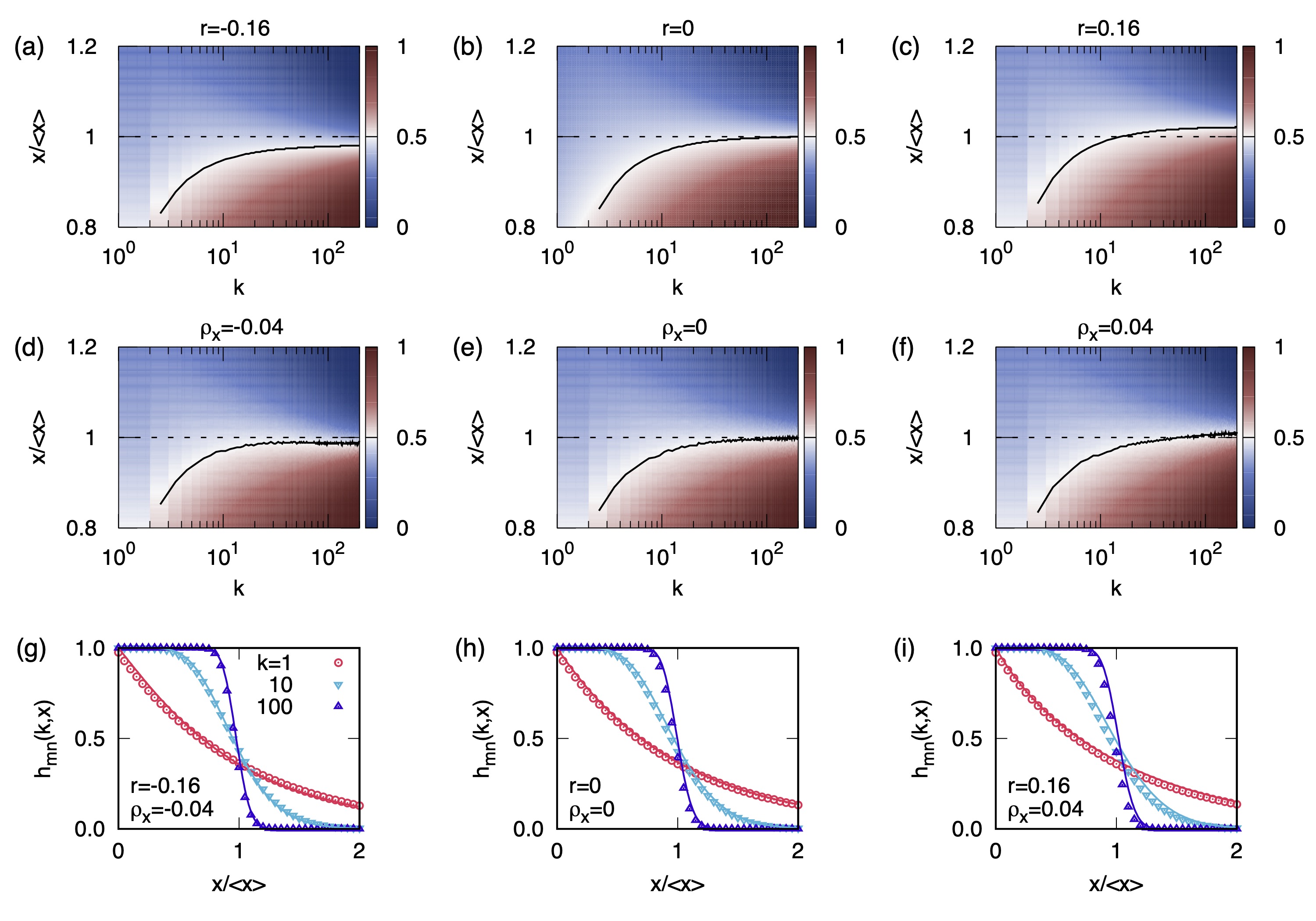}
\caption{Mean-based GFP: (a--c) Heat maps of the analytical solution of $h_{\rm mn}(k,x)$ in Eq.~\eqref{eq:hkx_mean_expo} in the cases with negatively correlated attributes ($r=-0.16$) (a), uncorrelated attributes ($r=0$) (b), and positively correlated attributes ($r=0.16$) (c). (d--f) Heat maps of the simulation results of $h_{\rm mn}(k,x)$ for $\rho_x=-0.04$ (d), $0$ (e), and $0.04$ (f). For each value of $\rho_x$, $2\times 10^3$ random configurations of attributes per network were generated using $P(x)$ with $\lambda=1$ in Eq.~\eqref{eq:expo_Px} for $10$ networks of size $N=5\times10^4$ with an exponential degree distribution with the average of $50$. In (a--f) solid lines denote $x^*_r(k)/\langle x\rangle$, where the transition point $x^*_r(k)$ is defined in Eq.~\eqref{eq:transition}, and horizontal dotted lines for $x/\langle x\rangle=1$ are included for guiding eyes. (g--i) Comparison of $h_{\rm mn}(k,x)$ for $k=1$, $10$, and $100$ between the analytical solution (solid lines) and simulation results (symbols) in the cases with $(r,\rho_x)=(-0.16,-0.04)$ (g), $(0,0)$ (h), and $(0.16,0.04)$ (i). Error bars for confidence intervals for simulation results at significance level $\alpha=0.05$ are smaller than symbols.}
\label{fig:mn}
\end{figure*}

By plugging Eq.~\eqref{eq:Pxx} into Eq.~\eqref{eq:joint_Px}, we obtain
\begin{align}
  & P(x_1,\cdots,x_k|x) = \prod_{j=1}^k P(x_j) \prod_{j=1}^k [1+rf(x_j)f(x)] \nonumber \\
  & \approx \prod_{j=1}^k P(x_j) \left[1+r\sum_{j=1}^k f(x_j)f(x)+\mathcal{O}(r^2) \right].
  \label{eq:joint_Px_expand}
\end{align}
For the second line, by assuming that $|r|\ll 1$, we have expanded the equation up to the first order of $r$. We take the Laplace transform of Eq.~\eqref{eq:mean_define} using Eq.~\eqref{eq:joint_Px_expand} to get
\begin{align}
  & \tilde h_{\rm mn}(k,s)\approx \frac{1}{s}\left[1-\tilde P\left(\frac{s}{k}\right)^k\right]   \nonumber \\ 
  & + rk\int_0^\infty dx_1 Q(x_1)
  \prod_{j=2}^k \int_0^\infty dx_j P(x_j) \int_0^{\bar x_k} dx e^{-sx} f(x) \nonumber \\
  & + \mathcal{O}(r^2),
  \label{eq:mean_laplace}
\end{align}
where 
\begin{align}
    Q(x)\equiv P(x)f(x),\ \bar x_k\equiv \frac{1}{k}\sum_{j=1}^k x_j
\end{align}
and $\tilde P(s)$ is the Laplace transform of $P(x)$. Then the mean-based peer pressure $h_{\rm mn}(k,x)$ can be obtained by taking the inverse Laplace transform of Eq.~\eqref{eq:mean_laplace} analytically or numerically if necessary. Note that the analytical result in Eq.~\eqref{eq:mean_laplace} has been derived for the arbitrary form of $P(x)$ and any correlation coefficient $\rho_{x}$ whose range is limited by the functional form of $P(x)$ [see Eq.~\eqref{eq:rhoxx_r}].

For studying the solvable case, we consider an exponential distribution for $x$, i.e.,
\begin{align}
  P(x)=\lambda e^{-\lambda x},
  \label{eq:expo_Px}
\end{align}
with $\langle x\rangle =1/\lambda$, leading to
\begin{align}
  Q(x)=\lambda e^{-\lambda x}-2\lambda e^{-2\lambda x}.
\end{align}
Note that $A=1/4$ from Eq.~\eqref{eq:rhoxx_r}, implying that 
\begin{align}
    \rho_x=\frac{r}{4}.
    \label{eq:rel_rhox_r}
\end{align}
Since 
\begin{align}
  \tilde P(s)=\frac{\lambda}{s+\lambda},\  \tilde Q(s)=\frac{-\lambda s}{(s+\lambda)(s+2\lambda)},
  \label{eq:expo_PsQs}
\end{align}
one gets from Eq.~\eqref{eq:mean_laplace}
\begin{align}
  &\tilde h_{\rm mn}(k,s) \approx \frac{1}{s}\left[1-\tilde P\left(\frac{s}{k}\right)^k\right] + rk \left[ -\frac{1}{s} \tilde P\left(\frac{s}{k}\right)^{k-1} \tilde Q\left(\frac{s}{k}\right) \right. \nonumber \\
  & \left. +\frac{2}{s+\lambda} \tilde P\left(\frac{s+\lambda}{k}\right)^{k-1} \tilde Q\left(\frac{s+\lambda}{k}\right) \right]
  +\mathcal{O}(r^2).
  \label{eq:mean_expo_laplace}
\end{align}
Then we take the inverse Laplace transform of Eq.~\eqref{eq:mean_expo_laplace} to finally get $h_{\rm mn}(k,x)$ as follows:
\begin{align}
  & h_{\rm mn}(k,x) \approx
  g(k,\lambda kx) + rk(-1)^{k+1} e^{-2\lambda kx} (1-2e^{-\lambda x})\nonumber \\ 
  & \times \left[ g(k,-\lambda kx)-1 \right]
  +\mathcal{O}(r^2),
  \label{eq:hkx_mean_expo}
\end{align}
where
\begin{align}
    g(a,z)\equiv \frac{\Gamma(a,z)}{\Gamma(a)},\ \Gamma(a,z)=\int_z^\infty t^{a-1}e^{-t}dt.
    \label{eq:gamma_fn}
\end{align}
Here $\Gamma(a)$ and $\Gamma(a,z)$ denote the Gamma function and the upper incomplete Gamma function, respectively. See Appendix~\ref{append:inv} for the detailed calculation. By this solution we can rigorously understand the effects of the correlation between attributes of neighboring nodes on the mean-based GFP. We also remark that the first term on the right hand side of Eq.~\eqref{eq:hkx_mean_expo} is the same as the previous analytical solution for the case with uncorrelated attributes, namely, Eq.~(10) with $\alpha=1$ in Ref.~\cite{Jo2014Generalized}. 

\begin{figure*}[!t]
\includegraphics[width=0.85\textwidth]{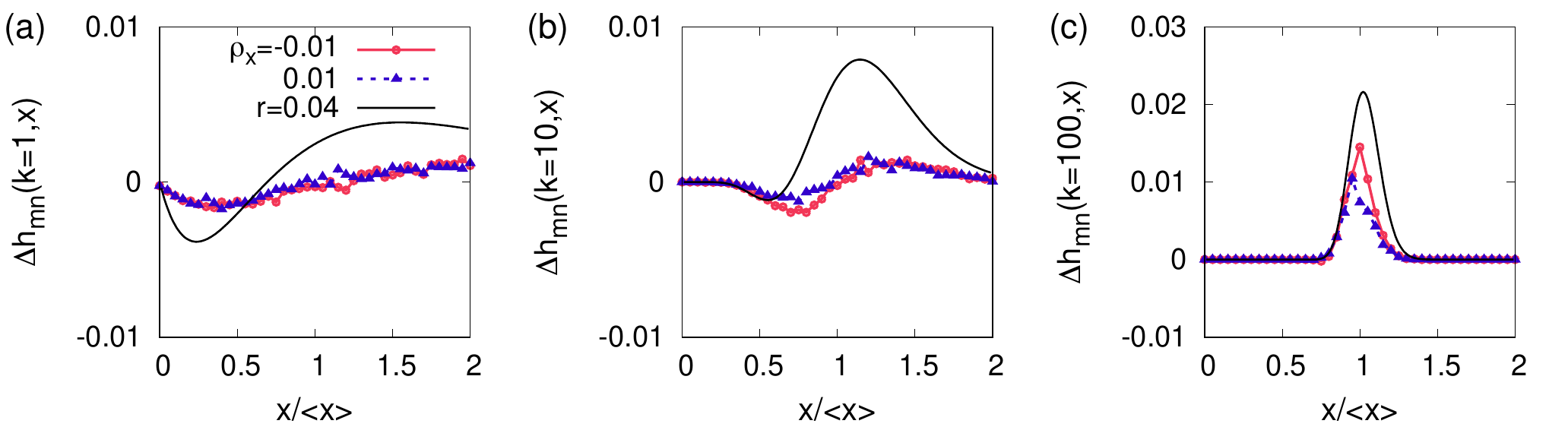}
\caption{Mean-based GFP: Differences in the peer pressure between the correlated cases ($\rho_x=-0.01$ and $0.01$) and the uncorrelated case ($\rho_x=0$), i.e., $\Delta h_{\rm mn}(k,x)$ in Eq.~\eqref{eq:dhkx_mn}, for $k=1$ (a), $10$ (b), and $100$ (c). For each value of $\rho_x$, $5\times 10^4$ random configurations of attributes per network were generated for $10$ networks used in Fig.~\ref{fig:mn}. In each panel, simulation results are denoted by symbols, while the analytical result, i.e., the first-order term of $r$ on the right hand side of Eq.~\eqref{eq:hkx_mean_expo} with $r=0.04$, is denoted by a solid line.}
\label{fig:mn_k}
\end{figure*}

The analytical solution in Eq.~\eqref{eq:hkx_mean_expo} is depicted as heat maps in Fig.~\ref{fig:mn}(a--c) for both uncorrelated and correlated cases, i.e., for $r=-0.16$, $0$, and $0.16$, respectively. The effects due to the attribute correlation can be effectively characterized by a \emph{transition point} $x^*_r(k)$ for given $k$ and $r$ between regimes with low and high peer pressure, which is defined by the condition 
\begin{align}
    h_{\rm mn}[k,x=x^*_r(k)]=\frac{1}{2}.
    \label{eq:transition}
\end{align}
The curves $x^*_r(k)$ for various values of $r$ are shown as solid lines in Fig.~\ref{fig:mn}(a--c). In particular, we find that as $k$ increases, $x^*_r(k)/\langle x\rangle$ approaches a value smaller than, equal to, or larger than one for $r<0$, $r=0$, or $r>0$, respectively. For the case with $r=0$ we indeed derive the following result from Eq.~\eqref{eq:hkx_mean_expo}:
\begin{align}
    \lim_{k\to\infty} g\left(k,k\tfrac{x}{\langle x\rangle}\right)=
    \begin{cases}
    1 & \textrm{if}\ 0\le \frac{x}{\langle x\rangle}<1,\\
    \frac{1}{2} & \textrm{if}\ \frac{x}{\langle x\rangle}=1,\\
    0 & \textrm{if}\ \frac{x}{\langle x\rangle}>1,
    \end{cases}
    \label{eq:g_largek}
\end{align}
see Appendix~\ref{append:g} for the derivation. These results imply that in a network in which attributes of neighboring nodes are positively correlated, some nodes having attributes above the average tend to have high peer pressure as their neighbors have even higher attributes than their own attributes, and vice versa. This behavior is qualitatively consistent with the previous numerical results presented in Fig. 2(c,~i) of Ref.~\cite{Jo2014Generalized}.

\subsection{Numerical simulation}\label{subsec:mean_numeric}

The analytical solution of $h_{\rm mn}(k,x)$ in Eq.~\eqref{eq:hkx_mean_expo} is compared with simulation results. We first generate $10$ uncorrelated random networks of size $N=5\times 10^4$ following the configuration model~\cite{Catanzaro2005Generation}, in which the degrees of nodes are drawn from an exponential distribution with the average of $50$. Each node $i$ in the generated network is assigned by an attribute $x_i$ that is randomly drawn from an exponential distribution $P(x)$ in Eq.~\eqref{eq:expo_Px}. For quantifying the correlation between attributes of neighboring nodes, we adopt the Pearson correlation coefficient, precisely,
\begin{align}
    \hat \rho_x\equiv \frac{L \sum_{ij} x_i x_j - [\sum_{ij} \frac{1}{2}(x_i + x_j)]^2}{L \sum_{ij} \frac{1}{2}(x_i^2+ x_j^2) - [\sum_{ij} \frac{1}{2}(x_i + x_j)]^2},
\end{align}
where the summations are over all links $ij$ and $L$ is the number of links in the network~\cite{Newman2003Mixinga}. To introduce the correlation between attributes of neighboring nodes, we uniformly randomly choose a link, say $ij$, and swap $x_i$ and $x_j$ only when the swap makes $\hat\rho_x$ closer to the target value $\rho_x$. This swapping is repeated until $\hat\rho_x$ gets close enough to $\rho_x$, i.e., until the following condition is satisfied:
\begin{align}
    |\hat\rho_x-\rho_x|<\epsilon
\end{align}
with $\epsilon=10^{-5}$. Once the network with (un)correlated attributes is ready, we calculate the mean-based peer pressure for each node $i$ by using
\begin{align}
h_{i,\rm mn}\equiv \theta\left(\frac{1}{k_i}\sum_{j\in\Lambda_i}x_j - x_i\right),
  \label{eq:hkx_mn_define}
\end{align}
to obtain the average of $h_{i,\rm mn}$ for nodes with the same $k$ and $x$, i.e., $h_{\rm mn}(k,x)$. For a given $\rho_x$, we generate $2\times 10^3$ random configurations of attributes per network. Since $\rho_x=r/4$ in Eq.~\eqref{eq:rel_rhox_r}, we obtain the simulation results for $\rho_x=-0.04$, $0$, and $0.04$ to compare them with the analytical counterparts for $r=-0.16$, $0$, and $0.16$, respectively. We find that the simulation results of $h_{\rm mn}(k,x)$ and of the transition point $x^*_r(k)$ in Fig.~\ref{fig:mn}(d--f) are comparable with the analytical solution in Fig.~\ref{fig:mn}(a--c). This is supported by the comparison between the simulation results and analytical solution of $h_{\rm mn}(k,x)$ as a function of $x$ for several values of $k$ in each case of $(\rho_x,r)=(-0.04,-0.16)$, $(0,0)$, and $(0.04,0.16)$, as shown in Fig.~\ref{fig:mn}(g--i).

Since the analytical solution of $h_{\rm mn}(k,x)$ in Eq.~\eqref{eq:hkx_mean_expo} was obtained up to the first-order term of $r$ by assuming that $|r|\ll 1$, the effects of higher-order terms of $r$ on the results can be studied by looking at the difference between the results in the correlated case and those in the uncorrelated case. Precisely, we compare the first-order term of $r$ on the right hand side of Eq.~\eqref{eq:hkx_mean_expo} for a given value of $r>0$ to the difference between correlated and uncorrelated cases, defined as
\begin{align}
 \Delta h_{\rm mn}(k,x)\equiv \begin{cases}
  h^{\rho_x}_{\rm mn}(k,x)-h^0_{\rm mn}(k,x) & \textrm{if}\ \rho_x>0,\\
  h^0_{\rm mn}(k,x)-h^{\rho_x}_{\rm mn}(k,x) & \textrm{if}\ \rho_x<0,
 \end{cases}
 \label{eq:dhkx_mn}
\end{align}
where $h^{\rho_x}_{\rm mn}(k,x)$ is the simulation result of the mean-based peer pressure when the correlation between attributes of neighboring nodes is given as $\rho_x$. Here $\rho_x=\pm r/4$ are used for the comparison. For the simulation, we generate $5\times 10^4$ random configurations of attributes per network. It is shown in Fig.~\ref{fig:mn_k} that the difference $\Delta h_{\rm mn}(k,x)$ is only qualitatively predicted by the first-order term of $r$ in Eq.~\eqref{eq:hkx_mean_expo}. This is probably due to the non-negligible higher-order terms of $r$, the calculation of which seems to be highly non-trivial. In addition, the deviation of the analytical solution from the simulation results might be partly due to the correlation between attributes of neighbors of the node. We have assumed the conditional independence between attributes of neighbors of the focal node $i$, i.e., $\{x_j\}_{j\in \Lambda_i}$, enabling us to simplify the joint PDF of those $x_j$s, as done in Eq.~\eqref{eq:joint_Px}. However, the correlation between $x_i$ and $x_j$ can naturally lead to the correlation between $x_j$s.

\begin{figure*}[!t]
\includegraphics[width=0.85\textwidth]{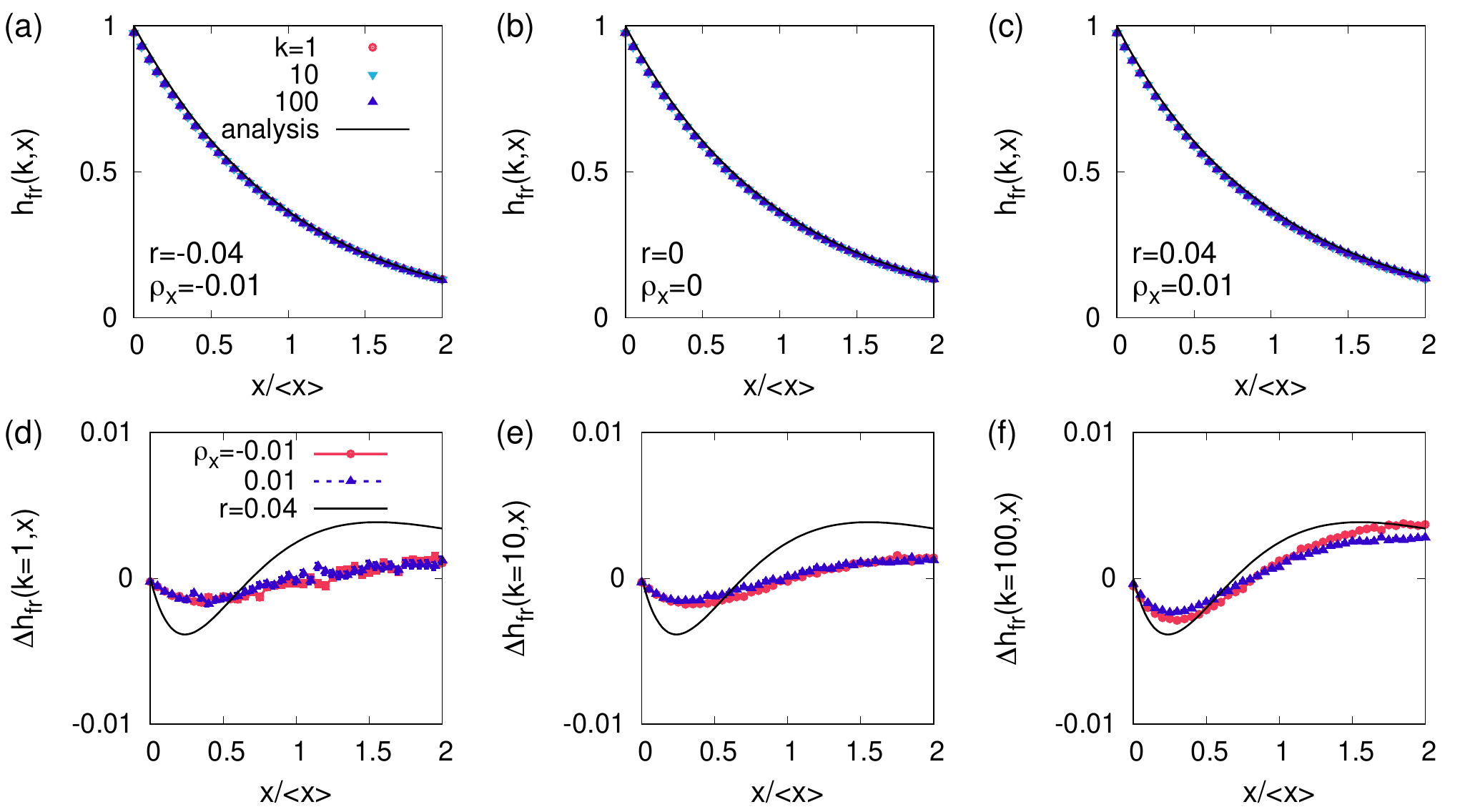}
\caption{Fraction-based GFP: (a--c) Comparison of the analytical solution of $h_{\rm fr}(k,x)$ in Eq.~\eqref{eq:hkx_fraction_expo} (solid lines) to simulation results (symbols) for $k=1$, $10$, and $100$ in the cases with $(r,\rho_x)=(-0.04,-0.01)$ (a), $(0,0)$ (b), and $(0.04,0.01)$ (c). The simulation results are obtained using the same set of random configurations of attributes on networks used in Fig.~\ref{fig:mn}. (d--f) Differences in the peer pressure between the correlated cases ($\rho_x=-0.01$ and $0.01$) and the uncorrelated case ($\rho_x=0$), i.e., $\Delta h_{\rm fr}(k,x)$, for $k=1$ (d), $10$ (e), and $100$ (f). Here $\Delta h_{\rm fr}(k,x)$ is defined similarly to $\Delta h_{\rm mn}(k,x)$ in Eq.~\eqref{eq:dhkx_mn}. The simulation results are obtained using the same set of random configurations of attributes on networks used in Fig.~\ref{fig:mn_k}. Simulation results are denoted by symbols, while the analytical result, i.e., the first-order term of $r$ on the right hand side of Eq.~\eqref{eq:hkx_fraction_expo} with $r=0.04$, is denoted by a solid line. In all panels, standard errors are smaller than symbols.}
\label{fig:fr}
\end{figure*}

\section{Fraction-based GFP}\label{sec:fraction}

Next we consider the fraction-based GFP in terms of the fraction of neighbors having higher attributes than the focal node~\cite{Lee2019Impact}. Here we define the fraction-based peer pressure as the ensemble average of the fraction of neighbors having higher attributes than the focal node with degree $k$ and attribute $x$ as follows:
\begin{align}
  & h_{\rm fr}(k,x) \equiv \bigg\langle \frac{1}{k}\sum_{j=1}^k \theta(x_j-x) \bigg\rangle \nonumber\\
  & =\prod_{j=1}^k \int_0^\infty dx_j P(x_1,\cdots,x_k|x) \frac{1}{k}\sum_{j=1}^k \theta(x_j-x).
  \label{eq:fraction_define}
\end{align}
By the assumption of conditional independence in Eq.~\eqref{eq:joint_Px} the fraction-based peer pressure in Eq.~\eqref{eq:fraction_define} reduces to a simpler form as 
\begin{align}
  h_{\rm fr}(k,x)=\int_0^\infty dx_1 P(x_1|x)\theta(x_1-x).
  \label{eq:fraction_Px}
\end{align}
It is remarkable that $h_{\rm fr}(k,x)$ is independent of $k$ and that from Eq.~\eqref{eq:mean_define}
\begin{align}
    h_{\rm fr}(k,x)=h_{\rm mn}(1,x).
    \label{eq:equiv_fr_mn}
\end{align}
Using Eq.~\eqref{eq:Pxx} one obtains
\begin{align}
  h_{\rm fr}(k,x)=\int_x^\infty dx_1 P(x_1)+ rf(x) \int_x^\infty dx_1 Q(x_1).
\end{align}
In the case with the exponential distribution of attributes in Eq.~\eqref{eq:expo_Px}, we have 
\begin{align}
  h_{\rm fr}(k,x)=e^{-\lambda x} + r(e^{-\lambda x} - 3e^{-2\lambda x} +2e^{-3\lambda x}).
  \label{eq:hkx_fraction_expo}
\end{align}
Note that this solution is exact without any assumption on the range of $r$.

The analytical solution of $h_{\rm fr}(k,x)$ in Eq.~\eqref{eq:hkx_fraction_expo} is depicted as solid lines in Fig.~\ref{fig:fr}(a--c) for both uncorrelated and correlated cases. As expected from the equivalence between the fraction-based GFP and the mean-based GFP for $k=1$ in Eq.~\eqref{eq:equiv_fr_mn}, we conclude that the positive correlation between attributes of neighboring nodes enhances (suppresses) the fraction-based peer pressure of nodes having higher (lower) attributes. The opposite behavior is found for the negatively correlated attributes.

For the comparison of the analytical solution to the simulation results, we use the same networks with correlated attributes generated for the mean-based GFP to measure the fraction-based peer pressure for each node $i$ by using
\begin{align}
h_{i,\rm fr}\equiv \frac{1}{k_i}\sum_{j\in\Lambda_i}\theta(x_j - x_i).
  \label{eq:hkx_fr_define}
\end{align}
Then we calculate the average of $h_{i,\rm fr}$ for nodes with the same $k$ and $x$ to get $h_{\rm fr}(k,x)$. The simulation results of $h_{\rm fr}(k,x)$ for various values of $\rho_x$ and $k$ are presented as symbols in Fig.~\ref{fig:fr}(a--c), which are comparable with the analytical solution. We also find in Fig.~\ref{fig:fr}(d--f) that the difference $\Delta h_{\rm fr}(k,x)$, defined similarly to $\Delta h_{\rm mn}(k,x)$ in Eq.~\eqref{eq:dhkx_mn}, is qualitatively predicted by the first-order term of $r$ in Eq.~\eqref{eq:hkx_fraction_expo}. In particular, when $k$ is large, e.g., $k=100$, the simulation results and the analytical solution are not only qualitatively but also quantitatively similar to each other. Yet the deviation between them might be attributed to the assumption on the conditional independence between attributes of neighbors of the focal node, as discussed in the previous Section.

\begin{figure*}[!t]
\includegraphics[width=0.9\textwidth]{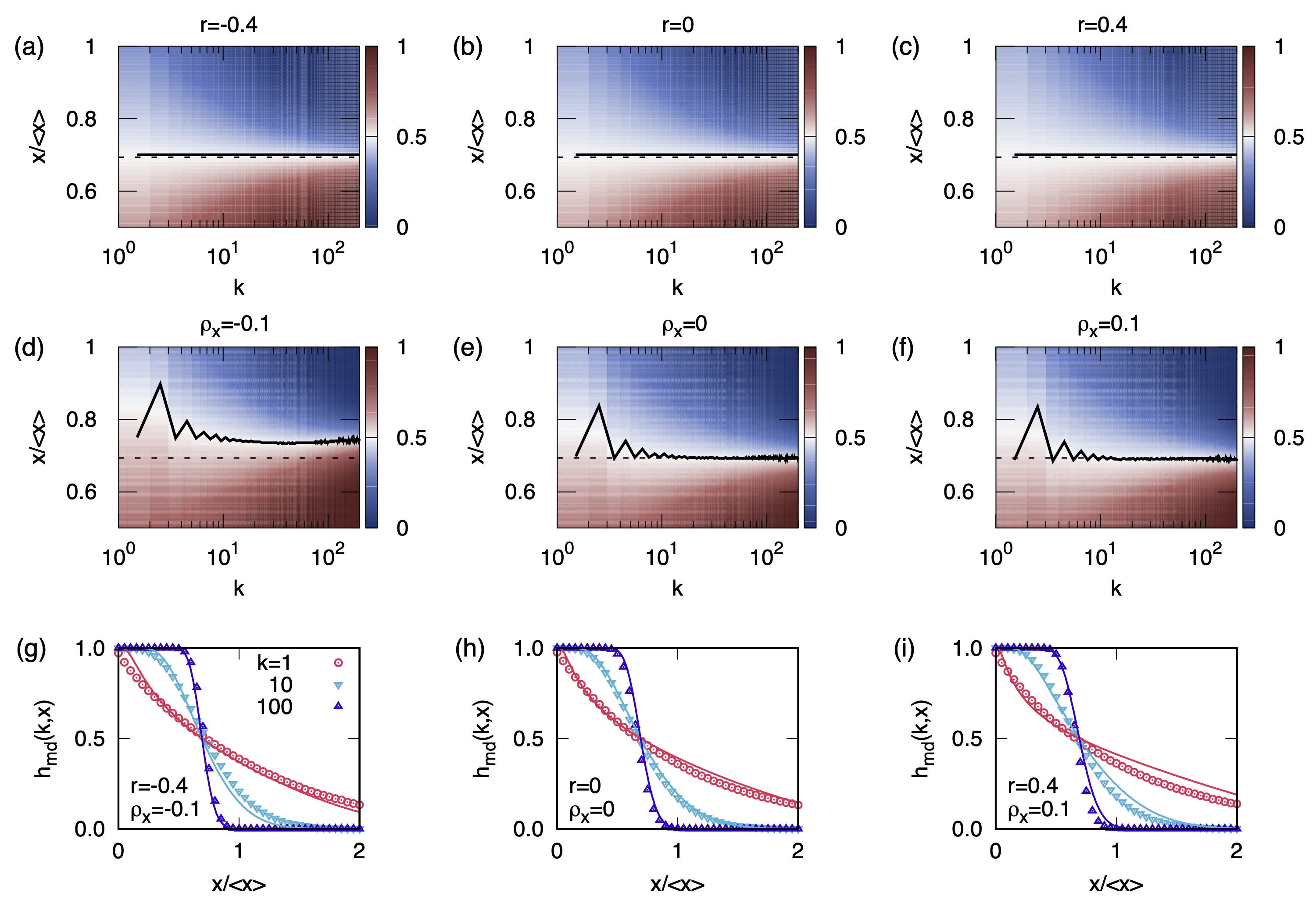}
\caption{Median-based GFP: (a--c) Heat maps of the analytical solution of $h_{\rm md}(k,x)$ in Eq.~\eqref{eq:hkx_median} for $r=-0.4$ (a), $0$ (b), and $0.4$ (c). (d--f) Heat maps of the simulation results of $h_{\rm md}(k,x)$ for $\rho_x=-0.1$ (d), $0$ (e), and $0.1$ (f). The simulation results are obtained using $2\times 10^3$ random configurations of attributes per network for $10$ networks used in Fig.~\ref{fig:mn}. In (a--f) solid lines denote $x^*_r(k)/\langle x\rangle$, where the transition point $x^*_r(k)$ is defined similarly to the mean-based case in Eq.~\eqref{eq:transition}, and horizontal dotted lines for $x/\langle x\rangle=\ln 2\approx 0.69$ are included for guiding eyes. (g--i) Comparison of $h_{\rm md}(k,x)$ for $k=1$, $10$, and $100$ between the analytical solution (solid lines) and simulation results (symbols) in the cases with $(r,\rho_x)=(-0.4,-0.1)$ (g), $(0,0)$ (h), and $(0.4,0.1)$ (i). Error bars for confidence intervals for simulation results at significance level $\alpha=0.05$ are smaller than symbols.}
\label{fig:md}
\end{figure*}

\section{Median-based GFP}\label{sec:median}

As mentioned in Section~\ref{sec:intro}, the median has been used instead of the average for summarizing the attributes of neighbors of the focal node~\cite{Feld1991Why, Ugander2011Anatomy} because the median is less sensitive to the neighbors whose attributes are very high, in particular when the attribute distribution $P(x)$ is skewed to the right. We define the median-based peer pressure as the probability that the focal node has lower attribute than the median attribute of its neighbors:
\begin{align}
  h_{\rm md}(k,x) \equiv \Bigg\langle \theta\left(\frac{1}{k}\sum_{j=1}^k \theta(x_j-x) -\frac{1}{2}\right) \Bigg\rangle.
  \label{eq:median_define}
\end{align}
Note that $h_{\rm md}(1,x)=h_{\rm fr}(k,x)=h_{\rm mn}(1,x)$. As the above equation in Eq.~\eqref{eq:median_define} is not trivial to analyze, by the assumption of conditional independence between $x_j$s, we rewrite the median-based peer pressure in terms of the binomial distribution:
\begin{align}
  h_{\rm md}(k,x) = \sum_{j=\lceil (k+1)/2 \rceil}^k {k \choose j} p_x^j(1-p_x)^{k-j},
  \label{eq:median_binom}
\end{align}
where $p_x$ is the probability that a neighbor of the focal node has an attribute bigger than that of the focal node, hence 
\begin{align}
    p_x\equiv \Pr(x_j>x)=h_{\rm fr}(k,x).    
\end{align}
A similar approach has been taken for the FP when the degrees of neighboring nodes are correlated~\cite{Wu2017NeighborNeighbor}. In the case with large $k$, the median-based peer pressure in Eq.~\eqref{eq:median_binom} can be approximated as
\begin{align}
  h_{\rm md}(k,x) \approx 1-\Phi\left[\frac{(1-2p_x)\sqrt{k}}{2\sqrt{p_x(1-p_x)}}\right],
  \label{eq:hkx_median}
\end{align}
where $\Phi(z)$ is the cumulative distribution function of the normal distribution:
\begin{align}
    \Phi(z)=\frac{1}{\sqrt{2\pi}}\int_{-\infty}^z e^{-t^2/2}dt.
\end{align}

For the exponentially distributed attributes, i.e., as in Eq.~\eqref{eq:expo_Px}, one can numerically calculate $h_{\rm md}(k,x)$ with $p_x=h_{\rm fr}(k,x)$ using Eq.~\eqref{eq:hkx_fraction_expo}. The analytical solution of $h_{\rm md}(k,x)$ in Eq.~\eqref{eq:hkx_median} is depicted as heat maps in Fig.~\ref{fig:md}(a--c) for both uncorrelated and correlated cases, i.e., $r=-0.4$, $0$, and $0.4$, respectively. We observe that $h_{\rm md}(k,x=\langle x\rangle\ln 2)=1/2$, irrespective of both $k$ and $r$, which can be easily shown by the fact that $p_x=1/2$ when $x=\langle x\rangle\ln 2$ from Eq.~\eqref{eq:hkx_fraction_expo}. That is, the transition point between regimes with low and high peer pressure, similarly defined as Eq.~\eqref{eq:transition}, is obtained as $x^*_r(k)=\langle x\rangle\ln 2$, which turns out to be constant of $r$ and $k$ [see also Fig.~\ref{fig:md}(a--c)]. The transition of $h_{\rm md}(k,x)$ at $x^*_r(k)$ becomes more gradual as the correlation increases. This implies that the positive correlation between attributes of neighboring nodes enhances (suppresses) the median-based peer pressure of nodes having higher (lower) attributes.

\begin{figure*}[!t]
\includegraphics[width=.85\textwidth]{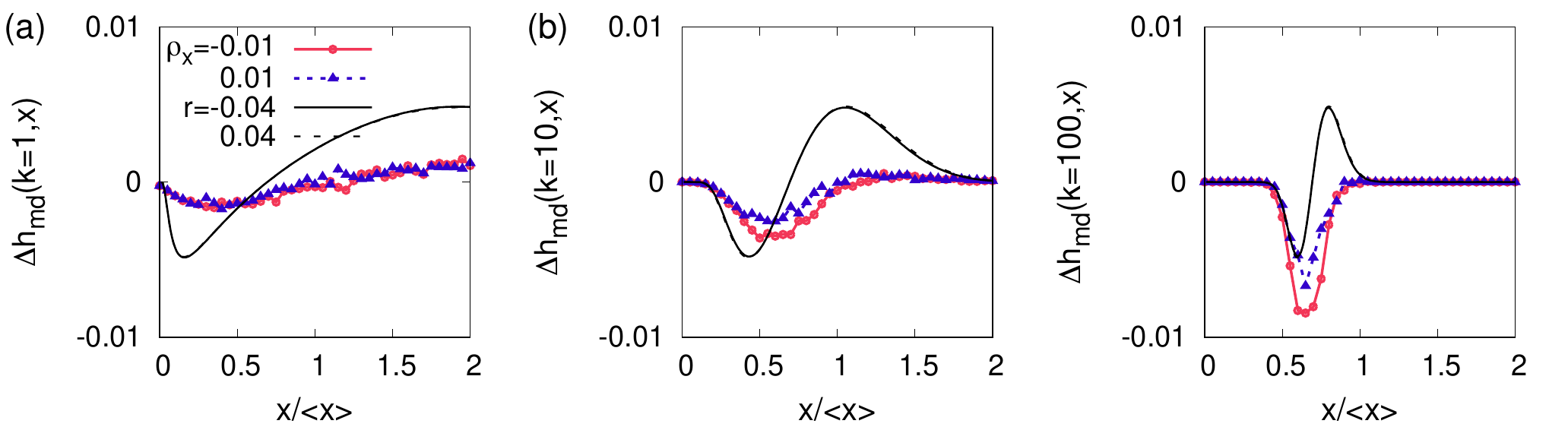}
\caption{Median-based GFP: Differences in the peer pressure between the correlated cases ($\rho_x=-0.01$ and $0.01$) and the uncorrelated case ($\rho_x=0$), i.e., $\Delta h_{\rm md}(k,x)$, for $k=1$ (a), $10$ (b), and $100$ (c). Here $\Delta h_{\rm md}(k,x)$ is defined similarly to $\Delta h_{\rm mn}(k,x)$ in Eq.~\eqref{eq:dhkx_mn}. The simulation results are obtained using $5\times 10^4$ random configurations of attributes per network for $10$ networks used in Fig.~\ref{fig:md}. Simulation results are denoted by symbols, while the analytical results derived from the right hand side of Eq.~\eqref{eq:hkx_median} with $r=\mp 0.04$, are denoted by solid lines and dashed lines, respectively.}
\label{fig:md_k}
\end{figure*}

We remark that for a given $r$, $x^*_r(k)$ is constant of $k$ for the median-based GFP, whereas it varies with $k$ for the mean-based GFP. It is because very large attributes of a node's neighbors significantly affect the average of neighbors' attributes but they may barely affect their median.

For the simulation, we generate the networks with correlated attributes similarly to the case of the mean-based GFP to measure the median-based peer pressure for each node $i$ by using
\begin{align}
h_{i,\rm md}\equiv \theta\left(x_{i,\rm md}-x_i\right),
  \label{eq:hkx_md_define}
\end{align}
where $x_{i,\rm md}$ is defined as the median of the set of attributes of $i$'s neighbors, i.e., $\{x_j\}_{j\in \Lambda_i}$. If the node $i$ has an even number of neighbors, $x_{i,\rm md}$ is given as the average of two middle attributes. Then we calculate the average of $h_{i,\rm md}$ for nodes with the same $k$ and $x$ to get $h_{\rm md}(k,x)$. The simulation results of $h_{\rm md}(k,x)$ for $\rho_x=-0.1$, $0$, and $0.1$ are shown in Fig.~\ref{fig:md}(d--f), respectively. We find that the simulation results for non-negatively correlated attributes are comparable with the analytical solution in Eq.~\eqref{eq:hkx_median}, whereas those for negatively correlated attributes deviate from the analytical solution, which is also confirmed in Fig.~\ref{fig:md}(g--i). The effect of correlations between attributes of neighboring nodes is also shown in terms of $\Delta h_{\rm md}(k,x)$, which is defined similarly to $\Delta h_{\rm mn}(k,x)$ in Eq.~\eqref{eq:dhkx_mn}. These differences are compared to the analytical counterparts that are derived from Eq.~\eqref{eq:hkx_median} in Fig.~\ref{fig:md_k} to find that the simulation results are only qualitatively explained by the analytical solution. The deviations between them are probably due to the assumption on the conditional independence as well as the assumption for the large $k$.

\section{Conclusion}\label{sec:concl}

The generalized friendship paradox (GFP) states that your neighbors have on average higher attributes than you do~\cite{Eom2014Generalized, Jo2014Generalized}. In other words, individuals tend to compare their own attributes to the average of attributes of their neighbors. However, the average is not the only summarization method of their neighborhood: The median of attributes of neighbors~\cite{Feld1991Why} as well as the fraction of neighbors having higher attributes than that of the individual~\cite{Lee2019Impact} have been used, but mostly for the friendship paradox. Yet little has been known about how such different summarization methods affect the GFP behavior for a network with correlated attributes. 

For the systematic comparison of different summarization methods, namely, mean-based, fraction-based, and median-based methods, we have derived the approximate analytical solutions of the probability that the GFP holds for an individual node with given degree and attribute, enabling us to understand the GFP more rigorously. For modeling the correlation between attributes of neighboring nodes, we have adopted a Farlie-Gumbel-Morgenstern (FGM) copula among others to successfully obtain the analytical solutions for the correlated cases; in the case of the mean-based method, the analytical solution is obtained only for weakly correlated attributes. These solutions are numerically confirmed and support some of the previous simulation results~\cite{Jo2014Generalized}, while we also find some systematic deviations at a finer scale between the analytical solutions and the simulation results. Such deviations might be partly due to the higher-order correlations as well as due to the correlation between attributes of neighbors of a focal node that has been ignored by the conditional independence between attributes of neighbors of the focal node for the sake of analytic tractability. These limits might be able to be overcome by considering the higher-order terms of the correlation parameter and the higher-order correlations between neighbors of the focal node. Finally, one can investigate the case with other functional forms of the attribute distribution than the exponential distribution, such as heavy-tailed ones~\cite{Clauset2009Powerlaw, Eom2014Generalized}, using other kinds of copulas~\cite{Nelsen2006Introduction} to represent more realistic correlation structure between attributes of neighboring nodes in complex networks.

\begin{acknowledgments}
H.-H.J. acknowledges financial support by Basic Science Research Program through the National Research Foundation of Korea (NRF) grant funded by the Ministry of Education (NRF-2018R1D1A1A09081919) and by the Catholic University  of Korea, Research Fund, 2020. This work was supported by the National Research Foundation of Korea (NRF) grant funded by the Korea government (Ministry of Science and ICT) (No. 2020R1G1A1101950).
\end{acknowledgments}

\appendix

\section{Inverse Laplace transform of Eq.~\eqref{eq:mean_expo_laplace}}\label{append:inv}

We use the following result for the Laplace transform:
\begin{align}
    \mathcal{L}\left\{\left[g(k,ax)-1\right]e^{-bx}\right\}(s)
    =-\frac{1}{s+b}\left(\frac{a}{s+a+b}\right)^k.
    \label{eq:invLaplace}
\end{align}
Using Eq.~\eqref{eq:expo_PsQs} the first term in the order of $r$ on the right hand side of Eq.~\eqref{eq:mean_expo_laplace} is explicitly written as
\begin{align}
    -\frac{1}{s} \tilde P\left(\frac{s}{k}\right)^{k-1} \tilde Q\left(\frac{s}{k}\right) = \frac{1}{s+2\lambda k}\left(\frac{\lambda k}{s+\lambda k}\right)^k.
    \label{eq:app1}
\end{align}
The inverse Laplace transform of Eq.~\eqref{eq:app1} is obtained using Eq.~\eqref{eq:invLaplace} by setting $a=-\lambda k$ and $b=2\lambda k$:
\begin{align}
    (-1)^{k+1}\left[g(k,-\lambda k x)-1\right]e^{-2\lambda kx}.
    \label{appeq:1st}
\end{align}
The second term in the order of $r$ on the right hand side of Eq.~\eqref{eq:mean_expo_laplace} is explicitly written as
\begin{align}
    &\frac{2}{s+\lambda} \tilde P\left(\frac{s+\lambda}{k}\right)^{k-1} \tilde Q\left(\frac{s+\lambda}{k}\right)\nonumber \\
    &= \frac{-2}{s+\lambda + 2\lambda k}\left(\frac{\lambda k}{s+\lambda+\lambda k}\right)^k.
    \label{eq:app2}
\end{align}
The inverse Laplace transform of Eq.~\eqref{eq:app2} is obtained using Eq.~\eqref{eq:invLaplace} by setting $a=-\lambda k$ and $b=\lambda + 2\lambda k$:
\begin{align}
    2 (-1)^{k}\left[g(k,-\lambda k x)-1\right]e^{-(\lambda+2\lambda k)x}.
    \label{appeq:2nd}
\end{align}
Combining Eqs.~\eqref{appeq:1st} and~\eqref{appeq:2nd} we finally get the first-order term of $r$ on the right hand side of Eq.~\eqref{eq:hkx_mean_expo}.

\section{Derivation of Eq.~\eqref{eq:g_largek}}\label{append:g}

Here we derive the results of $g(k,\lambda kx)$ in the limit of $k\to\infty$, i.e., Eq.~\eqref{eq:g_largek}. For the sake of simplicity, a new variable is defined as $u\equiv \lambda x=x/\langle x\rangle$, hence
\begin{align}
    g(k,\lambda kx)=\frac{\Gamma(k,uk)}{\Gamma(k)}=1-\frac{\gamma(k,uk)}{\Gamma(k)},
\end{align}
where $\gamma(\cdot,\cdot)$ is the lower incomplete Gamma function. By Eq.~5.11.3 in Ref.~\cite{Olver2020NIST} the asymptotic expansion of $\Gamma(k)$ for $k \to \infty$ up to the leading term is obtained as
\begin{align}
    \Gamma(k)\simeq e^{-k}k^{k-1/2} \sqrt{2\pi}.
\end{align}

We first consider the case with $u=1$, i.e., $x=\langle x\rangle$. The asymptotic expansion of $\Gamma(k,k)$ for $k \to \infty$ up to the leading term is written as (Eq.~8.11.12 in Ref.~\cite{Olver2020NIST})
\begin{align}
    \Gamma(k,k)\simeq e^{-k}k^{k-1/2} \sqrt{\frac{\pi}{2}},
\end{align}
leading to 
\begin{align}
    \lim_{k\to\infty} g(k,k)=\lim_{k\to\infty} \frac{\Gamma(k,k)}{\Gamma(k)}=\frac{1}{2}.
\end{align}
Next, when $u>1$ ($x>\langle x\rangle$), from Eq.~8.11.7 in Ref.~\cite{Olver2020NIST} we get the asymptotic expansion of $\Gamma(k,uk)$ as follows:
\begin{align}
    \Gamma(k,uk)\simeq \frac{e^{-uk}(uk)^k}{k(u-1)},
\end{align}
leading to 
\begin{align}
    \lim_{k\to\infty} g(k,uk)=\lim_{k\to\infty} \frac{e^{k(\ln u-u+1)}k^{-1/2}}{\sqrt{2\pi}(u-1)}=0,
\end{align}
where we have used the fact that $\ln u-u+1< 0$ for $u>1$. Finally, when $0\le u<1$ ($0\le x<\langle x\rangle$), from Eq.~8.11.6 in Ref.~\cite{Olver2020NIST} we get the asymptotic expansion of $\gamma(k,uk)$ as follows:
\begin{align}
    \gamma(k,uk)\simeq \frac{e^{-uk}(uk)^k}{k(1-u)},
\end{align}
leading to 
\begin{align}
    \lim_{k\to\infty} g(k,uk)=\lim_{k\to\infty} \left[1 - \frac{e^{k(\ln u-u+1)}k^{-1/2}}{\sqrt{2\pi}(1-u)}\right]=1,
\end{align}
where we have used the fact that $\ln u-u+1< 0$ for $0\le u<1$. Summarizing these results, we get Eq.~\eqref{eq:g_largek}.


\begin{thebibliography}{39}%
\makeatletter
\providecommand \@ifxundefined [1]{%
 \@ifx{#1\undefined}
}%
\providecommand \@ifnum [1]{%
 \ifnum #1\expandafter \@firstoftwo
 \else \expandafter \@secondoftwo
 \fi
}%
\providecommand \@ifx [1]{%
 \ifx #1\expandafter \@firstoftwo
 \else \expandafter \@secondoftwo
 \fi
}%
\providecommand \natexlab [1]{#1}%
\providecommand \enquote  [1]{``#1''}%
\providecommand \bibnamefont  [1]{#1}%
\providecommand \bibfnamefont [1]{#1}%
\providecommand \citenamefont [1]{#1}%
\providecommand \href@noop [0]{\@secondoftwo}%
\providecommand \href [0]{\begingroup \@sanitize@url \@href}%
\providecommand \@href[1]{\@@startlink{#1}\@@href}%
\providecommand \@@href[1]{\endgroup#1\@@endlink}%
\providecommand \@sanitize@url [0]{\catcode `\\12\catcode `\$12\catcode
  `\&12\catcode `\#12\catcode `\^12\catcode `\_12\catcode `\%12\relax}%
\providecommand \@@startlink[1]{}%
\providecommand \@@endlink[0]{}%
\providecommand \url  [0]{\begingroup\@sanitize@url \@url }%
\providecommand \@url [1]{\endgroup\@href {#1}{\urlprefix }}%
\providecommand \urlprefix  [0]{URL }%
\providecommand \Eprint [0]{\href }%
\providecommand \doibase [0]{https://doi.org/}%
\providecommand \selectlanguage [0]{\@gobble}%
\providecommand \bibinfo  [0]{\@secondoftwo}%
\providecommand \bibfield  [0]{\@secondoftwo}%
\providecommand \translation [1]{[#1]}%
\providecommand \BibitemOpen [0]{}%
\providecommand \bibitemStop [0]{}%
\providecommand \bibitemNoStop [0]{.\EOS\space}%
\providecommand \EOS [0]{\spacefactor3000\relax}%
\providecommand \BibitemShut  [1]{\csname bibitem#1\endcsname}%
\let\auto@bib@innerbib\@empty
\bibitem [{\citenamefont {Castellano}\ \emph {et~al.}(2009)\citenamefont
  {Castellano}, \citenamefont {Fortunato},\ and\ \citenamefont
  {Loreto}}]{Castellano2009Statistical}%
  \BibitemOpen
  \bibfield  {author} {\bibinfo {author} {\bibfnamefont {C.}~\bibnamefont
  {Castellano}}, \bibinfo {author} {\bibfnamefont {S.}~\bibnamefont
  {Fortunato}},\ and\ \bibinfo {author} {\bibfnamefont {V.}~\bibnamefont
  {Loreto}},\ }\bibfield  {title} {\bibinfo {title} {Statistical physics of
  social dynamics},\ }\href {https://doi.org/10.1103/revmodphys.81.591}
  {\bibfield  {journal} {\bibinfo  {journal} {Reviews of Modern Physics}\
  }\textbf {\bibinfo {volume} {81}},\ \bibinfo {pages} {591} (\bibinfo {year}
  {2009})}\BibitemShut {NoStop}%
\bibitem [{\citenamefont {Sen}\ and\ \citenamefont
  {Chakrabarti}(2014)}]{Sen2014Sociophysics}%
  \BibitemOpen
  \bibfield  {author} {\bibinfo {author} {\bibfnamefont {P.}~\bibnamefont
  {Sen}}\ and\ \bibinfo {author} {\bibfnamefont {B.~K.}\ \bibnamefont
  {Chakrabarti}},\ }\href@noop {} {\emph {\bibinfo {title} {Sociophysics: {{An
  Introduction}}}}}\ (\bibinfo  {publisher} {{Oxford University Press}},\
  \bibinfo {address} {{Oxford}},\ \bibinfo {year} {2014})\BibitemShut {NoStop}%
\bibitem [{\citenamefont {Albert}\ and\ \citenamefont
  {Barab{\'a}si}(2002)}]{Albert2002Statistical}%
  \BibitemOpen
  \bibfield  {author} {\bibinfo {author} {\bibfnamefont {R.}~\bibnamefont
  {Albert}}\ and\ \bibinfo {author} {\bibfnamefont {A.-L.}\ \bibnamefont
  {Barab{\'a}si}},\ }\bibfield  {title} {\bibinfo {title} {Statistical
  mechanics of complex networks},\ }\href
  {https://doi.org/10.1103/revmodphys.74.47} {\bibfield  {journal} {\bibinfo
  {journal} {Review of Modern Physics}\ }\textbf {\bibinfo {volume} {74}},\
  \bibinfo {pages} {47} (\bibinfo {year} {2002})}\BibitemShut {NoStop}%
\bibitem [{\citenamefont {Borgatti}\ \emph {et~al.}(2009)\citenamefont
  {Borgatti}, \citenamefont {Mehra}, \citenamefont {Brass},\ and\ \citenamefont
  {Labianca}}]{Borgatti2009Network}%
  \BibitemOpen
  \bibfield  {author} {\bibinfo {author} {\bibfnamefont {S.~P.}\ \bibnamefont
  {Borgatti}}, \bibinfo {author} {\bibfnamefont {A.}~\bibnamefont {Mehra}},
  \bibinfo {author} {\bibfnamefont {D.~J.}\ \bibnamefont {Brass}},\ and\
  \bibinfo {author} {\bibfnamefont {G.}~\bibnamefont {Labianca}},\ }\bibfield
  {title} {\bibinfo {title} {Network analysis in the social sciences},\ }\href
  {https://doi.org/10.1126/science.1165821} {\bibfield  {journal} {\bibinfo
  {journal} {Science}\ }\textbf {\bibinfo {volume} {323}},\ \bibinfo {pages}
  {892} (\bibinfo {year} {2009})}\BibitemShut {NoStop}%
\bibitem [{\citenamefont {Menczer}\ \emph {et~al.}(2020)\citenamefont
  {Menczer}, \citenamefont {Fortunato},\ and\ \citenamefont
  {Davis}}]{Menczer2020First}%
  \BibitemOpen
  \bibfield  {author} {\bibinfo {author} {\bibfnamefont {F.}~\bibnamefont
  {Menczer}}, \bibinfo {author} {\bibfnamefont {S.}~\bibnamefont {Fortunato}},\
  and\ \bibinfo {author} {\bibfnamefont {C.~A.}\ \bibnamefont {Davis}},\
  }\href@noop {} {\emph {\bibinfo {title} {A First Course in Network
  Science}}}\ (\bibinfo  {publisher} {{Cambridge University Press}},\ \bibinfo
  {address} {{Cambridge}},\ \bibinfo {year} {2020})\BibitemShut {NoStop}%
\bibitem [{\citenamefont {Jo}\ \emph {et~al.}(2018)\citenamefont {Jo},
  \citenamefont {Murase}, \citenamefont {T{\"o}r{\"o}k}, \citenamefont
  {Kert{\'e}sz},\ and\ \citenamefont {Kaski}}]{Jo2018Stylized}%
  \BibitemOpen
  \bibfield  {author} {\bibinfo {author} {\bibfnamefont {H.-H.}\ \bibnamefont
  {Jo}}, \bibinfo {author} {\bibfnamefont {Y.}~\bibnamefont {Murase}}, \bibinfo
  {author} {\bibfnamefont {J.}~\bibnamefont {T{\"o}r{\"o}k}}, \bibinfo {author}
  {\bibfnamefont {J.}~\bibnamefont {Kert{\'e}sz}},\ and\ \bibinfo {author}
  {\bibfnamefont {K.}~\bibnamefont {Kaski}},\ }\bibfield  {title} {\bibinfo
  {title} {Stylized facts in social networks: {{Community}}-based static
  modeling},\ }\href {https://doi.org/10.1016/j.physa.2018.02.023} {\bibfield
  {journal} {\bibinfo  {journal} {Physica A: Statistical Mechanics and its
  Applications}\ }\textbf {\bibinfo {volume} {500}},\ \bibinfo {pages} {23}
  (\bibinfo {year} {2018})}\BibitemShut {NoStop}%
\bibitem [{\citenamefont {Barab{\'a}si}\ and\ \citenamefont
  {Albert}(1999)}]{Barabasi1999Emergence}%
  \BibitemOpen
  \bibfield  {author} {\bibinfo {author} {\bibfnamefont {A.-L.}\ \bibnamefont
  {Barab{\'a}si}}\ and\ \bibinfo {author} {\bibfnamefont {R.}~\bibnamefont
  {Albert}},\ }\bibfield  {title} {\bibinfo {title} {Emergence of {{Scaling}}
  in {{Random Networks}}},\ }\href
  {https://doi.org/10.1126/science.286.5439.509} {\bibfield  {journal}
  {\bibinfo  {journal} {Science}\ }\textbf {\bibinfo {volume} {286}},\ \bibinfo
  {pages} {509} (\bibinfo {year} {1999})}\BibitemShut {NoStop}%
\bibitem [{\citenamefont {Broido}\ and\ \citenamefont
  {Clauset}(2019)}]{Broido2019Scalefree}%
  \BibitemOpen
  \bibfield  {author} {\bibinfo {author} {\bibfnamefont {A.~D.}\ \bibnamefont
  {Broido}}\ and\ \bibinfo {author} {\bibfnamefont {A.}~\bibnamefont
  {Clauset}},\ }\bibfield  {title} {\bibinfo {title} {Scale-free networks are
  rare},\ }\href {https://doi.org/10.1038/s41467-019-08746-5} {\bibfield
  {journal} {\bibinfo  {journal} {Nature Communications}\ }\textbf {\bibinfo
  {volume} {10}},\ \bibinfo {pages} {1017} (\bibinfo {year}
  {2019})}\BibitemShut {NoStop}%
\bibitem [{\citenamefont {Newman}(2002)}]{Newman2002Assortative}%
  \BibitemOpen
  \bibfield  {author} {\bibinfo {author} {\bibfnamefont {M.~E.~J.}\
  \bibnamefont {Newman}},\ }\bibfield  {title} {\bibinfo {title} {Assortative
  {{Mixing}} in {{Networks}}},\ }\href
  {https://doi.org/10.1103/physrevlett.89.208701} {\bibfield  {journal}
  {\bibinfo  {journal} {Physical Review Letters}\ }\textbf {\bibinfo {volume}
  {89}},\ \bibinfo {pages} {208701} (\bibinfo {year} {2002})}\BibitemShut
  {NoStop}%
\bibitem [{\citenamefont {Fortunato}(2010)}]{Fortunato2010Community}%
  \BibitemOpen
  \bibfield  {author} {\bibinfo {author} {\bibfnamefont {S.}~\bibnamefont
  {Fortunato}},\ }\bibfield  {title} {\bibinfo {title} {Community detection in
  graphs},\ }\href@noop {} {\bibfield  {journal} {\bibinfo  {journal} {Physics
  Reports}\ }\textbf {\bibinfo {volume} {486}},\ \bibinfo {pages} {75}
  (\bibinfo {year} {2010})}\BibitemShut {NoStop}%
\bibitem [{\citenamefont {{Pastor-Satorras}}\ \emph {et~al.}(2015)\citenamefont
  {{Pastor-Satorras}}, \citenamefont {Castellano}, \citenamefont
  {Van~Mieghem},\ and\ \citenamefont
  {Vespignani}}]{Pastor-Satorras2015Epidemic}%
  \BibitemOpen
  \bibfield  {author} {\bibinfo {author} {\bibfnamefont {R.}~\bibnamefont
  {{Pastor-Satorras}}}, \bibinfo {author} {\bibfnamefont {C.}~\bibnamefont
  {Castellano}}, \bibinfo {author} {\bibfnamefont {P.}~\bibnamefont
  {Van~Mieghem}},\ and\ \bibinfo {author} {\bibfnamefont {A.}~\bibnamefont
  {Vespignani}},\ }\bibfield  {title} {\bibinfo {title} {Epidemic processes in
  complex networks},\ }\href {https://doi.org/10.1103/RevModPhys.87.925}
  {\bibfield  {journal} {\bibinfo  {journal} {Reviews of Modern Physics}\
  }\textbf {\bibinfo {volume} {87}},\ \bibinfo {pages} {925} (\bibinfo {year}
  {2015})}\BibitemShut {NoStop}%
\bibitem [{\citenamefont {Masuda}\ \emph {et~al.}(2017)\citenamefont {Masuda},
  \citenamefont {Porter},\ and\ \citenamefont {Lambiotte}}]{Masuda2017Random}%
  \BibitemOpen
  \bibfield  {author} {\bibinfo {author} {\bibfnamefont {N.}~\bibnamefont
  {Masuda}}, \bibinfo {author} {\bibfnamefont {M.~A.}\ \bibnamefont {Porter}},\
  and\ \bibinfo {author} {\bibfnamefont {R.}~\bibnamefont {Lambiotte}},\
  }\bibfield  {title} {\bibinfo {title} {Random walks and diffusion on
  networks},\ }\href {https://doi.org/10.1016/j.physrep.2017.07.007} {\bibfield
   {journal} {\bibinfo  {journal} {Physics Reports}\ }\textbf {\bibinfo
  {volume} {716-717}},\ \bibinfo {pages} {1} (\bibinfo {year}
  {2017})}\BibitemShut {NoStop}%
\bibitem [{\citenamefont {Feld}(1991)}]{Feld1991Why}%
  \BibitemOpen
  \bibfield  {author} {\bibinfo {author} {\bibfnamefont {S.~L.}\ \bibnamefont
  {Feld}},\ }\bibfield  {title} {\bibinfo {title} {Why {{Your Friends Have More
  Friends Than You Do}}},\ }\href {https://doi.org/10.2307/2781907} {\bibfield
  {journal} {\bibinfo  {journal} {American Journal of Sociology}\ }\textbf
  {\bibinfo {volume} {96}},\ \bibinfo {pages} {1464} (\bibinfo {year}
  {1991})}\BibitemShut {NoStop}%
\bibitem [{\citenamefont {Grund}(2014)}]{Grund2014Why}%
  \BibitemOpen
  \bibfield  {author} {\bibinfo {author} {\bibfnamefont {T.}~\bibnamefont
  {Grund}},\ }\bibfield  {title} {\bibinfo {title} {Why {{Your Friends Are More
  Important}} and {{Special Than You Think}}},\ }\href
  {https://doi.org/10.15195/v1.a10} {\bibfield  {journal} {\bibinfo  {journal}
  {Sociological Science}\ }\textbf {\bibinfo {volume} {1}},\ \bibinfo {pages}
  {128} (\bibinfo {year} {2014})}\BibitemShut {NoStop}%
\bibitem [{\citenamefont {Higham}(2019)}]{Higham2019Centralityfriendship}%
  \BibitemOpen
  \bibfield  {author} {\bibinfo {author} {\bibfnamefont {D.~J.}\ \bibnamefont
  {Higham}},\ }\bibfield  {title} {\bibinfo {title} {Centrality-friendship
  paradoxes: When our friends are more important than us},\ }\href
  {https://doi.org/10.1093/comnet/cny029} {\bibfield  {journal} {\bibinfo
  {journal} {Journal of Complex Networks}\ }\textbf {\bibinfo {volume} {7}},\
  \bibinfo {pages} {515} (\bibinfo {year} {2019})}\BibitemShut {NoStop}%
\bibitem [{\citenamefont {Bollen}\ \emph {et~al.}(2017)\citenamefont {Bollen},
  \citenamefont {Gon{\c c}alves}, \citenamefont {{van de Leemput}},\ and\
  \citenamefont {Ruan}}]{Bollen2017Happiness}%
  \BibitemOpen
  \bibfield  {author} {\bibinfo {author} {\bibfnamefont {J.}~\bibnamefont
  {Bollen}}, \bibinfo {author} {\bibfnamefont {B.}~\bibnamefont {Gon{\c
  c}alves}}, \bibinfo {author} {\bibfnamefont {I.}~\bibnamefont {{van de
  Leemput}}},\ and\ \bibinfo {author} {\bibfnamefont {G.}~\bibnamefont
  {Ruan}},\ }\bibfield  {title} {\bibinfo {title} {The happiness paradox: Your
  friends are happier than you},\ }\href
  {https://doi.org/10.1140/epjds/s13688-017-0100-1} {\bibfield  {journal}
  {\bibinfo  {journal} {EPJ Data Science}\ }\textbf {\bibinfo {volume} {6}},\
  \bibinfo {pages} {4} (\bibinfo {year} {2017})}\BibitemShut {NoStop}%
\bibitem [{\citenamefont {Zhou}\ \emph {et~al.}(2020)\citenamefont {Zhou},
  \citenamefont {Jin},\ and\ \citenamefont {Zafarani}}]{Zhou2020Sentiment}%
  \BibitemOpen
  \bibfield  {author} {\bibinfo {author} {\bibfnamefont {X.}~\bibnamefont
  {Zhou}}, \bibinfo {author} {\bibfnamefont {S.}~\bibnamefont {Jin}},\ and\
  \bibinfo {author} {\bibfnamefont {R.}~\bibnamefont {Zafarani}},\ }\bibfield
  {title} {\bibinfo {title} {Sentiment {{Paradoxes}} in {{Social Networks}}:
  {{Why Your Friends Are More Positive Than You}}?},\ }\href@noop {} {\bibfield
   {journal} {\bibinfo  {journal} {Proceedings of the International AAAI
  Conference on Web and Social Media}\ }\textbf {\bibinfo {volume} {14}},\
  \bibinfo {pages} {798} (\bibinfo {year} {2020})}\BibitemShut {NoStop}%
\bibitem [{\citenamefont {Hodas}\ \emph {et~al.}(2013)\citenamefont {Hodas},
  \citenamefont {Kooti},\ and\ \citenamefont {Lerman}}]{Hodas2013Friendship}%
  \BibitemOpen
  \bibfield  {author} {\bibinfo {author} {\bibfnamefont {N.~O.}\ \bibnamefont
  {Hodas}}, \bibinfo {author} {\bibfnamefont {F.}~\bibnamefont {Kooti}},\ and\
  \bibinfo {author} {\bibfnamefont {K.}~\bibnamefont {Lerman}},\ }\bibfield
  {title} {\bibinfo {title} {Friendship {{Paradox Redux}}: {{Your Friends Are
  More Interesting Than You}}},\ }in\ \href@noop {} {\emph {\bibinfo
  {booktitle} {Proceedings of 7th {{International Conference}} on {{Weblogs}}
  and {{Social Media}}}}}\ (\bibinfo {year} {2013})\BibitemShut {NoStop}%
\bibitem [{\citenamefont {Eom}\ and\ \citenamefont
  {Jo}(2014)}]{Eom2014Generalized}%
  \BibitemOpen
  \bibfield  {author} {\bibinfo {author} {\bibfnamefont {Y.-H.}\ \bibnamefont
  {Eom}}\ and\ \bibinfo {author} {\bibfnamefont {H.-H.}\ \bibnamefont {Jo}},\
  }\bibfield  {title} {\bibinfo {title} {Generalized friendship paradox in
  complex networks: {{The}} case of scientific collaboration},\ }\href
  {https://doi.org/10.1038/srep04603} {\bibfield  {journal} {\bibinfo
  {journal} {Scientific Reports}\ }\textbf {\bibinfo {volume} {4}},\ \bibinfo
  {pages} {4603} (\bibinfo {year} {2014})}\BibitemShut {NoStop}%
\bibitem [{\citenamefont {Jo}\ and\ \citenamefont
  {Eom}(2014)}]{Jo2014Generalized}%
  \BibitemOpen
  \bibfield  {author} {\bibinfo {author} {\bibfnamefont {H.-H.}\ \bibnamefont
  {Jo}}\ and\ \bibinfo {author} {\bibfnamefont {Y.-H.}\ \bibnamefont {Eom}},\
  }\bibfield  {title} {\bibinfo {title} {Generalized friendship paradox in
  networks with tunable degree-attribute correlation},\ }\href
  {https://doi.org/10.1103/physreve.90.022809} {\bibfield  {journal} {\bibinfo
  {journal} {Physical Review E}\ }\textbf {\bibinfo {volume} {90}},\ \bibinfo
  {pages} {022809} (\bibinfo {year} {2014})}\BibitemShut {NoStop}%
\bibitem [{\citenamefont {Lerman}\ \emph {et~al.}(2016)\citenamefont {Lerman},
  \citenamefont {Yan},\ and\ \citenamefont {Wu}}]{Lerman2016Majority}%
  \BibitemOpen
  \bibfield  {author} {\bibinfo {author} {\bibfnamefont {K.}~\bibnamefont
  {Lerman}}, \bibinfo {author} {\bibfnamefont {X.}~\bibnamefont {Yan}},\ and\
  \bibinfo {author} {\bibfnamefont {X.-Z.}\ \bibnamefont {Wu}},\ }\bibfield
  {title} {\bibinfo {title} {The "{{Majority Illusion}}" in {{Social
  Networks}}},\ }\href {https://doi.org/10.1371/journal.pone.0147617}
  {\bibfield  {journal} {\bibinfo  {journal} {PLoS ONE}\ }\textbf {\bibinfo
  {volume} {11}},\ \bibinfo {pages} {e0147617} (\bibinfo {year}
  {2016})}\BibitemShut {NoStop}%
\bibitem [{\citenamefont {Momeni}\ and\ \citenamefont
  {Rabbat}(2016)}]{Momeni2016Qualities}%
  \BibitemOpen
  \bibfield  {author} {\bibinfo {author} {\bibfnamefont {N.}~\bibnamefont
  {Momeni}}\ and\ \bibinfo {author} {\bibfnamefont {M.}~\bibnamefont
  {Rabbat}},\ }\bibfield  {title} {\bibinfo {title} {Qualities and
  {{Inequalities}} in {{Online Social Networks}} through the {{Lens}} of the
  {{Generalized Friendship Paradox}}},\ }\href
  {https://doi.org/10.1371/journal.pone.0143633} {\bibfield  {journal}
  {\bibinfo  {journal} {PLoS ONE}\ }\textbf {\bibinfo {volume} {11}},\ \bibinfo
  {pages} {e0143633} (\bibinfo {year} {2016})}\BibitemShut {NoStop}%
\bibitem [{\citenamefont {Benevenuto}\ \emph {et~al.}(2016)\citenamefont
  {Benevenuto}, \citenamefont {Laender},\ and\ \citenamefont
  {Alves}}]{Benevenuto2016Hindex}%
  \BibitemOpen
  \bibfield  {author} {\bibinfo {author} {\bibfnamefont {F.}~\bibnamefont
  {Benevenuto}}, \bibinfo {author} {\bibfnamefont {A.~H.~F.}\ \bibnamefont
  {Laender}},\ and\ \bibinfo {author} {\bibfnamefont {B.~L.}\ \bibnamefont
  {Alves}},\ }\bibfield  {title} {\bibinfo {title} {The {{H}}-index paradox:
  Your coauthors have a higher {{H}}-index than you do},\ }\href
  {https://doi.org/10.1007/s11192-015-1776-2} {\bibfield  {journal} {\bibinfo
  {journal} {Scientometrics}\ }\textbf {\bibinfo {volume} {106}},\ \bibinfo
  {pages} {469} (\bibinfo {year} {2016})}\BibitemShut {NoStop}%
\bibitem [{\citenamefont {Alipourfard}\ \emph {et~al.}(2020)\citenamefont
  {Alipourfard}, \citenamefont {Nettasinghe}, \citenamefont {Abeliuk},
  \citenamefont {Krishnamurthy},\ and\ \citenamefont
  {Lerman}}]{Alipourfard2020Friendship}%
  \BibitemOpen
  \bibfield  {author} {\bibinfo {author} {\bibfnamefont {N.}~\bibnamefont
  {Alipourfard}}, \bibinfo {author} {\bibfnamefont {B.}~\bibnamefont
  {Nettasinghe}}, \bibinfo {author} {\bibfnamefont {A.}~\bibnamefont
  {Abeliuk}}, \bibinfo {author} {\bibfnamefont {V.}~\bibnamefont
  {Krishnamurthy}},\ and\ \bibinfo {author} {\bibfnamefont {K.}~\bibnamefont
  {Lerman}},\ }\bibfield  {title} {\bibinfo {title} {Friendship paradox biases
  perceptions in directed networks},\ }\href
  {https://doi.org/10.1038/s41467-020-14394-x} {\bibfield  {journal} {\bibinfo
  {journal} {Nature Communications}\ }\textbf {\bibinfo {volume} {11}},\
  \bibinfo {pages} {707} (\bibinfo {year} {2020})}\BibitemShut {NoStop}%
\bibitem [{\citenamefont {Fotouhi}\ \emph {et~al.}(2015)\citenamefont
  {Fotouhi}, \citenamefont {Momeni},\ and\ \citenamefont
  {Rabbat}}]{Fotouhi2015Generalized}%
  \BibitemOpen
  \bibfield  {author} {\bibinfo {author} {\bibfnamefont {B.}~\bibnamefont
  {Fotouhi}}, \bibinfo {author} {\bibfnamefont {N.}~\bibnamefont {Momeni}},\
  and\ \bibinfo {author} {\bibfnamefont {M.~G.}\ \bibnamefont {Rabbat}},\
  }\bibfield  {title} {\bibinfo {title} {Generalized {{Friendship Paradox}}:
  {{An Analytical Approach}}},\ }in\ \href
  {https://doi.org/10.1007/978-3-319-15168-7_43} {\emph {\bibinfo {booktitle}
  {Social {{Informatics}}}}},\ Vol.\ \bibinfo {volume} {8852},\ \bibinfo
  {editor} {edited by\ \bibinfo {editor} {\bibfnamefont {L.~M.}\ \bibnamefont
  {Aiello}}\ and\ \bibinfo {editor} {\bibfnamefont {D.}~\bibnamefont
  {McFarland}}}\ (\bibinfo  {publisher} {{Springer International Publishing}},\
  \bibinfo {address} {{Cham}},\ \bibinfo {year} {2015})\ pp.\ \bibinfo {pages}
  {339--352}\BibitemShut {NoStop}%
\bibitem [{\citenamefont {Ugander}\ \emph {et~al.}(2011)\citenamefont
  {Ugander}, \citenamefont {Karrer}, \citenamefont {Backstrom},\ and\
  \citenamefont {Marlow}}]{Ugander2011Anatomy}%
  \BibitemOpen
  \bibfield  {author} {\bibinfo {author} {\bibfnamefont {J.}~\bibnamefont
  {Ugander}}, \bibinfo {author} {\bibfnamefont {B.}~\bibnamefont {Karrer}},
  \bibinfo {author} {\bibfnamefont {L.}~\bibnamefont {Backstrom}},\ and\
  \bibinfo {author} {\bibfnamefont {C.}~\bibnamefont {Marlow}},\ }\href@noop {}
  {\bibinfo {title} {The {{Anatomy}} of the {{Facebook Social Graph}}}}
  (\bibinfo {year} {2011})\BibitemShut {NoStop}%
\bibitem [{\citenamefont {Kooti}\ \emph {et~al.}(2014)\citenamefont {Kooti},
  \citenamefont {Hodas},\ and\ \citenamefont {Lerman}}]{Kooti2014Network}%
  \BibitemOpen
  \bibfield  {author} {\bibinfo {author} {\bibfnamefont {F.}~\bibnamefont
  {Kooti}}, \bibinfo {author} {\bibfnamefont {N.~O.}\ \bibnamefont {Hodas}},\
  and\ \bibinfo {author} {\bibfnamefont {K.}~\bibnamefont {Lerman}},\
  }\bibfield  {title} {\bibinfo {title} {Network {{Weirdness}}: {{Exploring}}
  the {{Origins}} of {{Network Paradoxes}}},\ }in\ \href@noop {} {\emph
  {\bibinfo {booktitle} {Proceedings of the 8th {{International Conference}} on
  {{Weblogs}} and {{Social Media}}}}}\ (\bibinfo {year} {2014})\BibitemShut
  {NoStop}%
\bibitem [{\citenamefont {Lee}\ \emph {et~al.}(2019)\citenamefont {Lee},
  \citenamefont {Lee}, \citenamefont {Eom}, \citenamefont {Holme},\ and\
  \citenamefont {Jo}}]{Lee2019Impact}%
  \BibitemOpen
  \bibfield  {author} {\bibinfo {author} {\bibfnamefont {E.}~\bibnamefont
  {Lee}}, \bibinfo {author} {\bibfnamefont {S.}~\bibnamefont {Lee}}, \bibinfo
  {author} {\bibfnamefont {Y.-H.}\ \bibnamefont {Eom}}, \bibinfo {author}
  {\bibfnamefont {P.}~\bibnamefont {Holme}},\ and\ \bibinfo {author}
  {\bibfnamefont {H.-H.}\ \bibnamefont {Jo}},\ }\bibfield  {title} {\bibinfo
  {title} {Impact of perception models on friendship paradox and opinion
  formation},\ }\href {https://doi.org/10.1103/PhysRevE.99.052302} {\bibfield
  {journal} {\bibinfo  {journal} {Physical Review E}\ }\textbf {\bibinfo
  {volume} {99}},\ \bibinfo {pages} {052302} (\bibinfo {year}
  {2019})}\BibitemShut {NoStop}%
\bibitem [{\citenamefont {Nelsen}(2006)}]{Nelsen2006Introduction}%
  \BibitemOpen
  \bibfield  {author} {\bibinfo {author} {\bibfnamefont {R.~B.}\ \bibnamefont
  {Nelsen}},\ }\href {https://doi.org/10.1007/0-387-28678-0} {\emph {\bibinfo
  {title} {An {{Introduction}} to {{Copulas}}}}},\ Springer {{Series}} in
  {{Statistics}}\ (\bibinfo  {publisher} {{Springer New York}},\ \bibinfo
  {address} {{New York, NY}},\ \bibinfo {year} {2006})\BibitemShut {NoStop}%
\bibitem [{\citenamefont {Cossette}\ \emph {et~al.}(2013)\citenamefont
  {Cossette}, \citenamefont {C{\^o}t{\'e}}, \citenamefont {Marceau},\ and\
  \citenamefont {Moutanabbir}}]{Cossette2013Multivariate}%
  \BibitemOpen
  \bibfield  {author} {\bibinfo {author} {\bibfnamefont {H.}~\bibnamefont
  {Cossette}}, \bibinfo {author} {\bibfnamefont {M.-P.}\ \bibnamefont
  {C{\^o}t{\'e}}}, \bibinfo {author} {\bibfnamefont {E.}~\bibnamefont
  {Marceau}},\ and\ \bibinfo {author} {\bibfnamefont {K.}~\bibnamefont
  {Moutanabbir}},\ }\bibfield  {title} {\bibinfo {title} {Multivariate
  distribution defined with
  {{Farlie}}\textendash{{Gumbel}}\textendash{{Morgenstern}} copula and mixed
  {{Erlang}} marginals: {{Aggregation}} and capital allocation},\ }\href
  {https://doi.org/10.1016/j.insmatheco.2013.03.006} {\bibfield  {journal}
  {\bibinfo  {journal} {Insurance: Mathematics and Economics}\ }\textbf
  {\bibinfo {volume} {52}},\ \bibinfo {pages} {560} (\bibinfo {year}
  {2013})}\BibitemShut {NoStop}%
\bibitem [{\citenamefont {Schucany}\ \emph {et~al.}(1978)\citenamefont
  {Schucany}, \citenamefont {Parr},\ and\ \citenamefont
  {Boyer}}]{Schucany1978Correlation}%
  \BibitemOpen
  \bibfield  {author} {\bibinfo {author} {\bibfnamefont {W.~R.}\ \bibnamefont
  {Schucany}}, \bibinfo {author} {\bibfnamefont {W.~C.}\ \bibnamefont {Parr}},\
  and\ \bibinfo {author} {\bibfnamefont {J.~E.}\ \bibnamefont {Boyer}},\
  }\bibfield  {title} {\bibinfo {title} {Correlation {{Structure}} in
  {{Farlie}}-{{Gumbel}}-{{Morgenstern Distributions}}},\ }\href
  {https://doi.org/10.2307/2335922} {\bibfield  {journal} {\bibinfo  {journal}
  {Biometrika}\ }\textbf {\bibinfo {volume} {65}},\ \bibinfo {pages} {650}
  (\bibinfo {year} {1978})}\BibitemShut {NoStop}%
\bibitem [{\citenamefont {Takeuchi}(2010)}]{Takeuchi2010Constructing}%
  \BibitemOpen
  \bibfield  {author} {\bibinfo {author} {\bibfnamefont {T.~T.}\ \bibnamefont
  {Takeuchi}},\ }\bibfield  {title} {\bibinfo {title} {Constructing a bivariate
  distribution function with given marginals and correlation: Application to
  the galaxy luminosity function},\ }\href
  {https://doi.org/10.1111/j.1365-2966.2010.16778.x} {\bibfield  {journal}
  {\bibinfo  {journal} {Monthly Notices of the Royal Astronomical Society}\
  }\textbf {\bibinfo {volume} {406}},\ \bibinfo {pages} {1830} (\bibinfo {year}
  {2010})}\BibitemShut {NoStop}%
\bibitem [{\citenamefont {Jo}(2019)}]{Jo2019Analytically}%
  \BibitemOpen
  \bibfield  {author} {\bibinfo {author} {\bibfnamefont {H.-H.}\ \bibnamefont
  {Jo}},\ }\bibfield  {title} {\bibinfo {title} {Analytically solvable
  autocorrelation function for weakly correlated interevent times},\ }\href
  {https://doi.org/10.1103/PhysRevE.100.012306} {\bibfield  {journal} {\bibinfo
   {journal} {Physical Review E}\ }\textbf {\bibinfo {volume} {100}},\ \bibinfo
  {pages} {012306} (\bibinfo {year} {2019})}\BibitemShut {NoStop}%
\bibitem [{\citenamefont {Jo}\ \emph {et~al.}(2019)\citenamefont {Jo},
  \citenamefont {Lee}, \citenamefont {Hiraoka},\ and\ \citenamefont
  {Jung}}]{Jo2019Copulabased}%
  \BibitemOpen
  \bibfield  {author} {\bibinfo {author} {\bibfnamefont {H.-H.}\ \bibnamefont
  {Jo}}, \bibinfo {author} {\bibfnamefont {B.-H.}\ \bibnamefont {Lee}},
  \bibinfo {author} {\bibfnamefont {T.}~\bibnamefont {Hiraoka}},\ and\ \bibinfo
  {author} {\bibfnamefont {W.-S.}\ \bibnamefont {Jung}},\ }\bibfield  {title}
  {\bibinfo {title} {Copula-based algorithm for generating bursty time
  series},\ }\href {https://doi.org/10.1103/PhysRevE.100.022307} {\bibfield
  {journal} {\bibinfo  {journal} {Physical Review E}\ }\textbf {\bibinfo
  {volume} {100}},\ \bibinfo {pages} {022307} (\bibinfo {year}
  {2019})}\BibitemShut {NoStop}%
\bibitem [{\citenamefont {Catanzaro}\ \emph {et~al.}(2005)\citenamefont
  {Catanzaro}, \citenamefont {Bogu{\~n}{\'a}},\ and\ \citenamefont
  {{Pastor-Satorras}}}]{Catanzaro2005Generation}%
  \BibitemOpen
  \bibfield  {author} {\bibinfo {author} {\bibfnamefont {M.}~\bibnamefont
  {Catanzaro}}, \bibinfo {author} {\bibfnamefont {M.}~\bibnamefont
  {Bogu{\~n}{\'a}}},\ and\ \bibinfo {author} {\bibfnamefont {R.}~\bibnamefont
  {{Pastor-Satorras}}},\ }\bibfield  {title} {\bibinfo {title} {Generation of
  uncorrelated random scale-free networks},\ }\href
  {https://doi.org/10.1103/physreve.71.027103} {\bibfield  {journal} {\bibinfo
  {journal} {Physical Review E}\ }\textbf {\bibinfo {volume} {71}},\ \bibinfo
  {pages} {027103} (\bibinfo {year} {2005})}\BibitemShut {NoStop}%
\bibitem [{\citenamefont {Newman}(2003)}]{Newman2003Mixinga}%
  \BibitemOpen
  \bibfield  {author} {\bibinfo {author} {\bibfnamefont {M.~E.~J.}\
  \bibnamefont {Newman}},\ }\bibfield  {title} {\bibinfo {title} {Mixing
  patterns in networks},\ }\href {https://doi.org/10.1103/physreve.67.026126}
  {\bibfield  {journal} {\bibinfo  {journal} {Physical Review E}\ }\textbf
  {\bibinfo {volume} {67}},\ \bibinfo {pages} {026126} (\bibinfo {year}
  {2003})}\BibitemShut {NoStop}%
\bibitem [{\citenamefont {Wu}\ \emph {et~al.}(2017)\citenamefont {Wu},
  \citenamefont {Percus},\ and\ \citenamefont
  {Lerman}}]{Wu2017NeighborNeighbor}%
  \BibitemOpen
  \bibfield  {author} {\bibinfo {author} {\bibfnamefont {X.-Z.}\ \bibnamefont
  {Wu}}, \bibinfo {author} {\bibfnamefont {A.~G.}\ \bibnamefont {Percus}},\
  and\ \bibinfo {author} {\bibfnamefont {K.}~\bibnamefont {Lerman}},\
  }\bibfield  {title} {\bibinfo {title} {Neighbor-{{Neighbor Correlations
  Explain Measurement Bias}} in {{Networks}}},\ }\href
  {https://doi.org/10.1038/s41598-017-06042-0} {\bibfield  {journal} {\bibinfo
  {journal} {Scientific Reports}\ }\textbf {\bibinfo {volume} {7}},\ \bibinfo
  {pages} {5576} (\bibinfo {year} {2017})}\BibitemShut {NoStop}%
\bibitem [{\citenamefont {Clauset}\ \emph {et~al.}(2009)\citenamefont
  {Clauset}, \citenamefont {Shalizi},\ and\ \citenamefont
  {Newman}}]{Clauset2009Powerlaw}%
  \BibitemOpen
  \bibfield  {author} {\bibinfo {author} {\bibfnamefont {A.}~\bibnamefont
  {Clauset}}, \bibinfo {author} {\bibfnamefont {C.~R.}\ \bibnamefont
  {Shalizi}},\ and\ \bibinfo {author} {\bibfnamefont {M.~E.~J.}\ \bibnamefont
  {Newman}},\ }\bibfield  {title} {\bibinfo {title} {Power-law distributions in
  empirical data},\ }\href {https://doi.org/10.1137/070710111} {\bibfield
  {journal} {\bibinfo  {journal} {SIAM Review}\ }\textbf {\bibinfo {volume}
  {51}},\ \bibinfo {pages} {661} (\bibinfo {year} {2009})}\BibitemShut
  {NoStop}%
\bibitem [{Olv(2 15)}]{Olver2020NIST}%
  \BibitemOpen
  \href@noop {} {\bibinfo {title} {{{NIST Digital Library}} of {{Mathematical
  Functions}}}},\ \bibinfo {howpublished} {http://dlmf.nist.gov/} (\bibinfo
  {year} {Release 1.1.0 of 2020-12-15})\BibitemShut {NoStop}%
\end{thebibliography}

%

\end{document}